\documentclass[authoryear]{elsarticle}
\usepackage{amsmath,amssymb}
\usepackage{times}
\usepackage{graphicx}
\usepackage{color}
\usepackage[T1]{fontenc}

\begin{document}

\title{Quantum noise may limit the mechanosensory sensitivity of cilia in the left--right organizer of the vertebrate bodyplan}

\author{Julyan H. E. Cartwright}
\affiliation{Instituto Andaluz de Ciencias de la Tierra, CSIC--Universidad de Granada, 18100 Armilla, Granada, Spain}
\affiliation{Instituto Carlos I de F{\'{\i}}sica Te{\'o}rica y Computacional,  Universidad de Granada,  18071 Granada, Spain}

\begin{abstract}
Could nature be harnessing quantum mechanics in cilia to optimize the sensitivity of the  mechanism of left--right symmetry breaking during development in vertebrates? I evaluate whether mechanosensing ---
 i.e., the detection of a left-right asymmetric signal through mechanical stimulation of sensory cilia, as opposed to biochemical signalling ---
might be functioning in the embryonic left--right organizer of the vertebrate bodyplan through quantum mechanics.  
I conclude that there is a possible role for quantum biology in mechanosensing in cilia.
The system may not be limited by classical thermal noise, but instead by quantum noise, with an amplification process providing active cooling. 

 \end{abstract}


\maketitle

\section{What's the limit to the sensitivity of mechanosensing cilia?}

Cilia possess enormously sensitive mechanosensory capabilities. This is apparent, for example, from the minute sounds that both we, and other species, are capable of detecting at the threshold of hearing \citep{majka2015}. A further example of these capabilities may be found in the cilia involved in the determination of left and right in the development of the body plan of many vertebrate species \citep{cartwright2004}. There, cilia both break the left--right symmetry of the embryo during the development of the body plan, by stirring a liquid to produce a leftward flow, and also detect that broken symmetry. This symmetry breaking during development is what leads to our hearts being on the left and our livers, on the right of our bodies.  Although it is now well established how motile cilia produce the leftward flow, it is still not understood how the flow is sensed.
One of the mechanisms currently being investigated by which the broken symmetry may be detected is  mechanosensing. 
However, it is not clear how  mechanosensing would function in the case of left--right symmetry breaking because the signal would be  weak, similar in magnitude to the noise, so that the signal to noise ratio would be rather poor \citep{ferreira2017,cartwright2020}.

 A similar question was discussed in the 1980s with regard to cilia in the auditory system. One interesting proposal made then was that quantum effects might be of importance in the ear. 
Bialek and colleagues argued that the sensitivity of the auditory system could be limited by quantum \citep{clerk2010}, rather than classical noise \citep{bialek1984,bialek1985}.
Although there is nothing wrong with the physical argument,
that proposal was withdrawn, at least partially, after a re-evaluation of the operating parameters of the cilia involved in hearing \citep{bialek1986}. Following this, further work in the late 1980s suggested that the ear might be functioning near to the classical thermal noise limit \citep{denk1989}. 

However,  although auditory cilia might not utilize quantum mechanics, perhaps other cilia may.  
Since the 1980s,  a number of examples have been identified  in which quantum mechanics may be of importance in biology, and quantum biology has begun to emerge as a field \citep{huelga2013,mcfadden2018,marais2018,cao2020}. 
Here I examine whether mechanosensory cilia may be making use of quantum mechanics to maximize their signal to noise ratio, and, in particular, whether this physics could be relevant to  mechanosensory cilia involved in left--right determination in vertebrate developmental biology in the biosystem known as the embryonic organizer.

\section{Left--right symmetry breaking in the embryonic organizer}

The organizer is the name given to an organ that appears transitorily  in the vertebrate embryo during organismal development. It appears to provide the earliest signal leading to left--right symmetry breaking in many biological model organisms, which, like ourselves, develop to be approximately bilaterally symmetric on the outside, but break that symmetry on the inside \citep{dasgupta2016}. We humans have our hearts on the left and livers on the right, for instance, in 9999 out of 10000 people. The organizer in many organisms contains motile cilia \citep{essner2002}.
In the mouse, where it is termed the node, the organizer is a shallow liquid-filled cavity some tens of micrometres across, stirred by whirling monocilia \citep{nonaka1998,okada1999}. (Monocilia, also called primary cilia, are a type of cilium constituted with a particular so-called $9+0$ structural arrangement of microtubules.) And in the zebrafish, where it is termed Kupffer's vesicle (KV), it is a spherical  liquid-filled cavity likewise stirred by monocilia \citep{essner2005,kramer-zucker2005}. 

It is now well established that, in these two instances at least, it is fluid mechanics that biology is using to break left--right symmetry \citep{cartwright2009}. A  flow set up across the organizer by the cilia whirling in a given sense stirs the fluid,
the chiral whirl and posterior tilt  together produce left--right asymmetric flow, 
so that the flow indicates which side is left \citep{cartwright2004,cartwright2008fluid,cartwright2008}. 
What is still not established, however, is how this fluid flow is detected by  biological systems in order to initiate  symmetry breaking; that is to say, the cascade of asymmetric left--right gene expression \citep{cartwright2007,cartwright2020}. 
Possible mechanisms are some form of chemosensing, i.e., detecting the presence of a given molecule in the fluid, or mechanosensing; detecting the magnitude or direction of the flow. 

 \begin{figure}[b]
\centering \includegraphics[width=\columnwidth,clip=true]{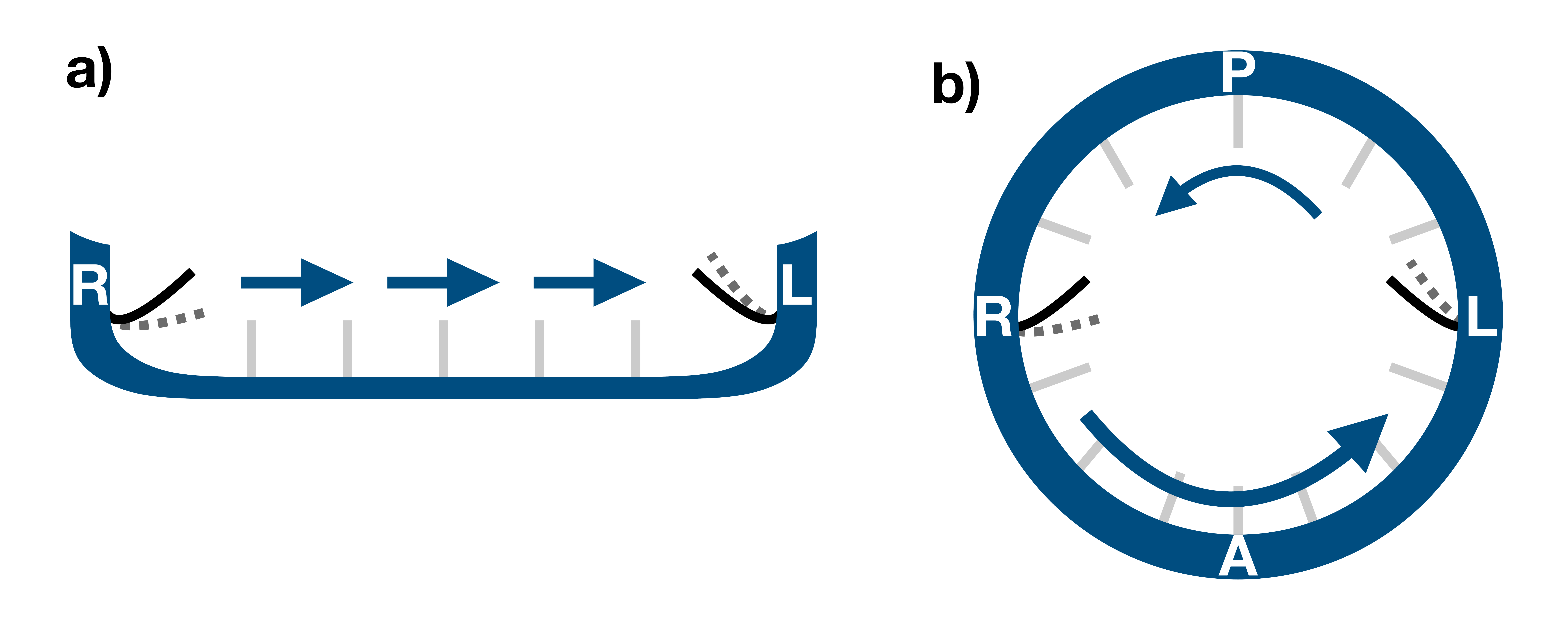}
 \caption{Mechanosensing mechanism as it is proposed to function in  a) the mouse node and b) the zebrafish Kupffer's vesicle. Owing to their finite size, cilia from the left/right side of the node or KV could potentially bend differentially (grey dotted lines) when exposed to a left-right asymmetric cilia-driven flow (arrows).  From \cite{cartwright2020}.
  \label{fig:mechanosensing}}
 \end{figure}

A chemosensory mechanism would seem achievable from the point of view of fluid physics \citep{cartwright2004,cartwright2008}, but to date no molecule involved has been identified. 
A mechanosensory mechanism, depicted in Fig.~\ref{fig:mechanosensing}, on the other hand, which, 
given this lack of a candidate morphogen molecule for chemosensing, might be considered more plausible biologically, 
presents puzzling problems to explain how it might function in terms of physics \citep{cartwright2020}. 
The question is that the signal is very small compared to the assumed noise, so that such a mechanosensory system would seem hardly reliable, yet failures of the left--right organizer system are remarkably rare in nature. In humans, as we have mentioned, only approximately one in ten thousand people has \emph{situs inversus}, in which the internal organs are the mirror image of the normal configuration. 

\section{The problem of the signal to noise ratio}

In the zebrafish KV, the signal threshold has been estimated  in terms of the torque at the base of a cilium. \cite{ferreira2017} noted the threshold torque averaging the torques detected on  three cilia on one side of the vesicle at $10^{-19}$~Nm, which they estimated to correspond to a shear rate of 0.5~s$^{-1}$, or a shear stress of 0.5~mPa, for a 6~$\mu$m long cilium. They compared this threshold with that of renal cilia, which are known to be mechanosensory, where the  shear stress  leading to a signal has been experimentally measured at 20~mPa  \citep{rydholm2010}, 1--2 orders of magnitude higher.
Equivalent data  for the mouse, in which normal wild-type mouse embryos have 200--300 motile cilia,  have not yet been obtained.
But \cite{shinohara2012} reported that mutant mice embryos with only two motile cilia  nevertheless still produce consistently elevated gene expression on the right, implying that these embryos have correctly established the left--right axis, despite very weak and localized flow.  
This implies that the signal threshold in these mice is likely lower than the above estimate for the KV. 

As a comparison, in the ear,  an estimate of signal threshold obtained by  \cite{bialek1985} is 
$10^{-18}$ W.  This is for an oscillator with angular frequency $\omega = 10^3$ rad s$^{-1}$ --- a typical auditory angular frequency --- and 
power ${\displaystyle P={\boldsymbol {\tau }}\cdot {\boldsymbol {\omega }},}$
so the associated threshold torque is $\tau = 10^{-21}$ Nm, two orders of magnitude lower than the zebrafish KV estimate of Ferreira et al.
In the mutant mouse embryos mentioned above, the attenuation of flow with distance from the two motile cilia
 --- at least $\mathcal{O}(r^{-1})$ for a Stokeslet,  $\mathcal{O}(r^{-2})$ for a rotlet, which are idealized models in microscale fluid dynamics of the flow due to a concentrated force \citep{blake1974} --- implies that this diminution must be taken into account.  In fact, owing to boundary effects at the ciliated surface the Stokeslet and rotlet will decay at least one order faster, i.e., $\mathcal{O}(r^{-2})$  for Stokeslet and $\mathcal{O}(r^{-3})$  for rotlet. 
 If the flow is rotlet dominated, as  with just two motile cilia, 
flow attenuation  with distance would imply a threshold torque a thousand times lower if the mechanosensory cilia are
at a distance of ten ciliary radii away from the motile cilia at the periphery of the nodal cavity,  for instance.
 That is to say, the results of  \cite{shinohara2012} mean that we must bear in mind a lower estimate, similar to that for the auditory system, as a baseline to the organizer sensitivity in the mouse.

We may compare these numbers with evaluations of the classical noise in these systems. 
\cite{bialek1985} estimated the effective fluid displacement noise in the ear as $\sim 1.5 \times 10^{-11}$ m; 1.5 times ---  the same order of magnitude as  --- the signal, approximately $10^{-18}$ W or $10^{-21}$ Nm.
\cite{ferreira2017} calculated for thermal noise on an elastic cilium in the KV  
a torque measured at the base of the cilium of $2.5\times 10^{-19}$ Nm. This noise torque corresponds  to a tip deflection of the torque divided by $KL = 3 El/L^2$, where $L = 6$~$\mu$m and $El =  3\times 10^{-23}$~Nm$^2$, so that the tip deflection should be 1 $\mu$m.
As the cilia in the murine organizer are similar to these, 
an equivalent estimation for the mouse should be of similar magnitude.
Thus we find that signal levels are lower than classical noise levels in the  organizer system under the supposition of mechanosensing.

\section{Can quantum biology come to the rescue?}

 \begin{figure}[tb]
\centering \includegraphics[width=\columnwidth,clip=true]{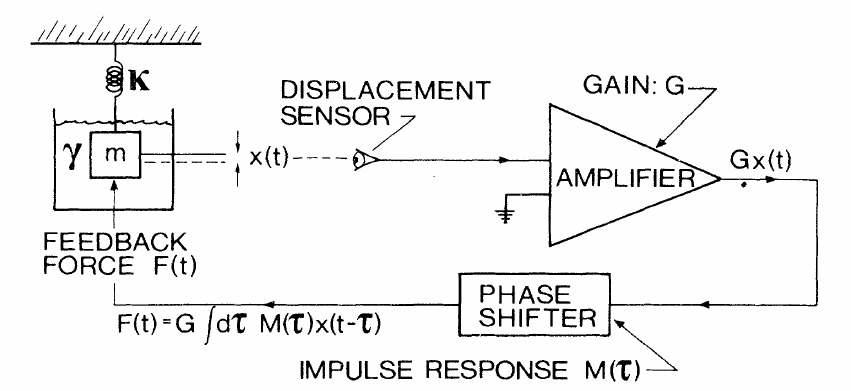}
 \caption{A sketch \citep{bialek1984} of the feedback amplification mechanism proposed by Gold in 1948 \citep{gold1948} to hold a mass-spring system out of equilibrium and later proposed by  \cite{bialek1985} to account for how the ear might be quantum-noise limited. Similarly, the organizer might be utilizing this type of mechanism.
 \label{fig:gold}}
 \end{figure}
 
Clearly this low signal to noise ratio represents a challenge for understanding the physical basis of the mechanosensing hypothesis in the organizer.
So might the noise levels in fact be lower than this classical limit? Could measurement in this system be governed by  quantum mechanics? 
Gold in the 1940s conceived that a biological ciliary system --- the ear, once again --- might use active mechanical elements  \citep{gold1948}. The  concept,  sketched in Fig.~\ref{fig:gold}, is that feedback from an amplifier should hold a cilium, a mass--spring system, out of equilibrium.  Later,  in the 1980s Bialek proposed   that the ear might use a system of this type in a state  from which thermal noise is eliminated, leaving only quantum noise  \citep{bialek1984}. 
Might such an active feedback  process  be operating in the cilia of 
the organizer?

Under certain conditions such a nonequilibrium system at biological body temperature could exhibit the same noise characteristics as at the temperature of absolute zero, where
 classical thermal noise is zero and the only source of noise is quantum. 
 If Fig.~\ref{fig:gold} is relevant to our system,  the amplifier used in the feedback process must have the minimum noise temperature allowed by quantum mechanics, so that the system is limited by quantum noise.
\cite{caves1982} and  \cite{bialek1983}  demonstrated that such an amplifier must add noise to any signal it processes, and this added noise must be at least the equivalent of doubling the zero-point noise associated with the input signal.

If feedback shifts the resonance in frequency by a large amount, then the force noise added by the feedback amplifier is approximately \citep{braginsky1992} 
$$
F_Q^2 = \hbar  (m \kappa)^{1/2}\omega^2 = \kappa \hbar \omega =  \kappa \hbar (\kappa/m)^{1/2},
$$
where $m$ is the mass, $\kappa$ is a spring constant  --- so $\omega$ is the natural frequency --- and $\hbar$ the reduced Planck constant. Since fluid displacements apply forces $F\sim \gamma \omega x_{fl}$ to the cilium, we can also put this noise in terms of a length scale $x_{fl}$ of fluid displacement.
To apply this to the organizer, direct measurement of the spring constant of a primary cilium --- albeit not a motile one: the measurement was of an immotile monocilium from a different biosystem, the canine kidney --- has given a value of $\kappa=7\times 10^{-5}$~N/m \citep{flaherty2020}. The cilium  may be estimated as having a mass of $4\times 10^{-17}$~kg  \citep{bialek1985}. An estimate of the quantum noise level is then $3\times 10^{-17}$~N, giving a maximum associated torque for a 6~$\mu$m long cilium of $2\times 10^{-22}$~Nm.

\section{Not in the ear, but in the left--right organizer}

Bialek and Schweitzer had used a value of the spring constant two orders of magnitude greater,  $8\times 10^{-3}$~N/m, in their first work on the auditory system \citep{bialek1985} and one order of magnitude greater,  $10^{-4}$~N/m, in their later work \citep{bialek1986}. In those cases, however, the estimates are for a different type of cilia,  for $9+2$ stereocilia, which might be assumed to be stiffer given that they have an extra central pair of microtubules over the $9+0$ monocilia without the central pair found in the organizer system.
With the first value, they gave the force noise $F_Q\sim 10^{-17}$~N, or $\sim 1.5\times 10^{-11}$~m in terms of displacement. That would indicate that  there is not much difference between the lower bound on the quantum-limited signal and the upper bound on the threshold signal that is reliably detected. That is,  the ear detects signals close to the quantum noise limit, and hence  hearing is a macroscopic quantum phenomenon. However, after this initial suggestion that cilia in the auditory system might utilize quantum mechanics, Bielek and Schweitzer in part withdrew their 1985 proposal  following a re-evaluation of the parameters of the cilia involved in the auditory system. They determined  \citep{bialek1986}  the effective quantum noise in terms of fluid displacement in the ear to be $\sim 5\times 10^{-14}$ m, substantially smaller than their estimates of the threshold displacement.
\cite{bialek2012} adjudged that he had failed to show that the quantum limits to measurement that we have discussed could be relevant to that biological system.
  \cite{ferreira2017}  concluded that the ear was close to the thermal noise limit, but detecting stimuli at higher frequencies and therefore acting as high- or band-pass filters.
 
To return to our system, unlike in the ear, in the mouse at least, based on  the findings of  \cite{shinohara2012}, the organizer is below the thermal noise limit and close to the quantum noise limit.  It is, moreover,  known that the primary cilium can  biochemically regulate its stiffness \citep{nguyen2015}.  In principle, then, a cell could use this variable stiffness to regulate its mechanosensing apparatus.  Thus although hearing might not depend on the quantum limits to measurement, it is possible that the organizer, if it depends on mechanosensing,  may be optimized in sensitivity by harnessing quantum mechanics.  
It should prove fruitful to compare the monocilium as a mechanosensor with mechanical resonators that approach the quantum limit in position-detection sensitivity or that have been cooled to a low resonator temperature \citep{poot2012}.

\section{So mechanosensing it is, then?}

Exactly how cilium displacement being coupled to molecules out of equilibrium and with quantum-mechanical coherence might be achieved chemically and biologically remains to be seen, but an important theoretical point from the side of physics is that it would enable a mechanosensing mechanism to be viable.  And from the position it was possible to take up until recently that the question of the sensory mechanism operating in the organizer is still an open one \citep{cartwright2020}, and despite work indicating that 
primary cilia are not calcium-responsive mechanosensors \citep{delling2016}, the latest experimental results in both 
 zebrafish \citep{sampaio2022} and mouse \citep{katoh2022} lead one seemingly inexorably towards the conclusion that the organizer system must be using mechanosensing.

There is an additional aspect  worth noting in this concept of feedback amplification that may possibly explain another observation  in the organizer system. An active system such as that described may become unstable so that it spontaneously oscillates. In the ear, this instability manifests itself as spontaneous otoacoustic emission, when the ear emits sounds. In the case of the organizer, some cilia seem to be motile, others not. Following this observation, it has been proposed that there are two distinct populations of cilia in the mouse node \citep{mcgrath2003}. But if the cilia in the organizer are utilizing feedback amplification, 
these immotile and motile cilia might not in fact be  two different types of cilia, but rather those that are subthreshold, are immotile and  act as detectors, and those that are superthreshold,  are motile and act as stirrers.

\section*{Declaration of competing interest}

The author has no competing interests.

\section*{Acknowledgements}

Over the years, I have had many valuable discussions with my colleagues Oreste Piro and Idan Tuval about the organizer system, about the auditory system, about biological fluid flow, and about ciliary dynamics; I gratefully acknowledge their inputs into the ideas I have put forward here. I acknowledge the contribution of the COST Action CA21169, Dynalife, supported by COST (European Cooperation in Science and Technology).

\bibliographystyle{elsarticle-harv}
\bibliography{cilia}

\end{document}